\documentclass[aip,amsmath,amssymb,reprint]{revtex4-1}

\usepackage{graphicx}
\usepackage{dcolumn}
\usepackage{bm}
\usepackage{color,soul}
\usepackage[utf8]{inputenc}
\usepackage[T1]{fontenc}
\usepackage{mathptmx}
\usepackage{mathtools}
\usepackage[font=small,labelfont=bf,
   justification=raggedright,
   format=plain]{caption}
\usepackage{subcaption}

\begin{document}

\preprint{AIP/123-QED}

\title{Computation of the equilibrium three-particle entropy for dense atomic fluids by molecular dynamics simulation}
\affiliation{ 
Department of Mathematics, Swinburne University of Technology, PO Box 218, Hawthorn, Victoria 3122, Australia
}

\author{Luca Maffioli}
\author{Nathan Clisby}
\author{Federico Frascoli}
\author{B. D. Todd}
  \email{btodd@swin.edu.au}


\begin{abstract}
We have computed the two and three-particle contribution to the entropy of a Weeks-Chandler-Andersen fluid via molecular dynamics simulations. The three-particle correlation function and entropy were computed with a new method which simplified calculation. Results are qualitatively similar to Lennard-Jones systems. We observed a numerical instability in the three-particle contribution. This phenomenon has been previously detected when the traditional method is used, thus it is likely to be intrinsic in the computation. While the effect of statistical fluctuations can be removed through an extrapolation procedure, the discretization error due to finite bin size is more difficult to characterize. With a correct choice of the bin size, a good estimate of the three-particle entropy contribution can be achieved at any state, even close to the freezing point. We observed that, despite the fact that the magnitude of the three-particle contribution increases significantly compared to the two-particle contribution as freezing is approached, the error induced from overestimation of the excess entropy by the two and three-body terms exceeds that induced by approximating the excess entropy with the two body term alone.
\end{abstract}

\maketitle

\section{\label{sec:intsec}Introduction}

Gibbs' entropy for a system of $N$ particles in thermodynamic
equilibrium at temperature $T$ is defined as \cite{green}
\begin{equation}
\label{eqn:gibbs}
S=-k_B\int{f(\bm{\Gamma}) \text{ln} f(\bm{\Gamma}) \text{d} \bm{\Gamma}}
\end{equation}   
with $f(\bm{\Gamma})$ the probability density function of the system in
the phase space point $\bm{\Gamma}$ and $k_B$ the Boltzmann constant.
There are several way to compute the entropy of a system via computer
simulation, and a direct estimation of the quantity in
Eq.~(\ref{eqn:gibbs}) can be considered the most direct way to achieve
it. This is, however, unfeasible for systems of more than a few
particles, due to the high dimensionality of the density function $f$.
Two methods have been found to compute the entropy of a system of
interacting particles via computer simulations. The first\cite{frenkel}
is based on the perturbation of the inter-particle potential from a
reference system, in which the entropy can be computed exactly (in the
case of a fluid, this system is the perfect gas), to the system of
interest. An alternative method is based on the Green's expansion
\cite{green}, in which the total entropy of a system of $N$ particles is
expressed as a sum of $N$ terms
\begin{equation}
\label{eqn:green}
S=S_1+S_2+...+S_N
\end{equation}   
with a generic term $S_i$ representing the contribution to the total
entropy of the $i$-body correlation. In this work, we will use the
second method, as we focus, in the first place, on the computation of
the three-body correlation function, and secondly, on the analysis of
the contribution of each term to the total entropy of a system.

Defining $f_N^{(i)}$ as the probability density function of a generic subset of $i$ particles (with $i\le N$), each term $S_i$ can be expressed as
\begin{equation}
\label{eqn:green's}
S_i=-\frac{k_B}{i!}\int f_N^{(i)}(1,...,i)\text{ln}(\delta f_N^{(i)}(1,...,i))\text{d} \bm{\Gamma_1}...\text{d}\bm{\Gamma_i},
\end{equation}
with
\begin{equation}
\delta f_N^{(i)}(1,...,i)=\frac{f_N^{(i)}(1,...,i)}{\prod_{k=1}^{i}f_N^{(i-1)}(\{1,...,i\} \setminus\{k\})}.
\end{equation}

Here the denominator in the second term of Eq.~(\ref{eqn:green's})
represents the product of the correlation function of all possible
subsets that can be formed removing one single element from the set
$\{1,...,i\}$. 

In cases of low density the first two terms of the expansion $S_1$ and
$S_2$, which can be computed straightforwardly via computer simulations,
are able to detect up to $90\%$ of the total entropy \cite{barevans2}.

Eq.~(\ref{eqn:green's}) has been variously simplified, by first
exploiting the fact that, for Canonical and Grand Canonical Ensembles,
$f^{(i)}$ can be factorized into a product of momentum and
configurational density functions \cite{wallace}
\begin{equation}
\label{eqn:factor}
f_N^{(i)}=f_N^{(1)}(\textbf{p}_1)...f_N^{(1)}(\textbf{p}_i)g_N^{(i)}(\textbf{r}_1,...,\textbf{r}_i).
\end{equation}
A local formulation (i.e. ensemble invariant) has been found \cite{nettletongreen,raveche,barevans1},
\begin{equation}
\label{eqn:entropylocalapp}
\begin{split}
\frac{S}{Nk_B}&=s_{pg}-\frac{\rho}{2}\int {g}^{(2)}\ln( {g}^{(2)})d\textbf{r}+\frac{\rho}{2}\int\biggl({g}^{(2)}-1\biggr)d\textbf{r} \\
&-\frac{\rho^2}{6}\int \biggl( {g}^{(3)}\ln( \delta{g}^{(3)})\biggr)d\textbf{r}^2 \\
&+\frac{\rho^2}{6}\int \biggl(  {g}^{(3)}-3{g}^{(2)}{g}^{(2)}+3{g}^{(2)}-1\biggr)d\textbf{r}^2+...
\end{split}
\end{equation}
Eq.~(\ref{eqn:entropylocalapp}) represents the first three terms of the
local formulation of the Green's expansion. 	$\rho$ is the density of
the system and $s_{pg}$ is the entropy of a perfect gas. It takes into
account  all information about momenta and about the configurational
distribution function $g^{(N)}$ of a perfect gas in the same macroscopic
state. We note that $s_{pg}$, whose expression is \cite{barevans1}
\begin{equation}
\label{eqn:spg}
s_{pg}=\frac{5}{2}-\ln\biggl[\rho\biggl(\frac{h}{(2\pi m k_B T)^{1/2}}\biggr)^3\biggr]
\end{equation}
with $m$ the mass of each particle, $h$ Planck's constant, does not
correspond to $S_1$ in Eq.~(\ref{eqn:green's})\cite{barevans1}; however
such considerations are secondary for the purpose of this work. For the
sake of clarity, the subscript $N$ will be omitted from now on. 

Eq.~(\ref{eqn:entropylocalapp}) has been used to compute the entropy of
a system of Lennard-Jones particles in various macroscopic conditions,
and has been shown to be able to detect no less than $70\%$ of the total
entropy \cite{barevans2}. However, computation of $s_3$ is affected by a
severe numerical instability, which manifests itself as an unphysical
drift for large distances, and makes the estimate of the three-body
entropy challenging. This behavior has been detected in previous
works\cite{barevans2,barevans3}, and correctly attributed to numerical
issues, although the limited computational resources available at the
time have not allowed a systematic investigation of it. Furthermore, the
computation of the three-particle function $g^{(3)}$ is not trivial, and
requires care.

In this work we first present a simpler method for the computation of
the three-particle function. Secondly, the two and thee-particle entropy
is computed for WCA\cite{wca} particles, the causes of this drift are
analyzed, and a method to correct it is presented.

\section{\label{sec:nmethsec}The dimensionless method}
\subsection{\label{sec:tradmeth}The traditional method}

At equilibrium, two and three-body distribution functions, $g^{(2)}$ and
$g^{(3)}$, depend respectively only on the distance between particles,
$r$, $s$, $t$. With a small abuse of notation
\begin{equation}
\label{eqn:eqfunctions}
\begin{split}
{g}^{(2)}(\textbf{r}_1,\textbf{r}_2)&={g}^{(2)}(r)\\
{g}^{(3)}(\textbf{r}_1,\textbf{r}_2,\textbf{r}_3)&={g}^{(3)}(r,s,t)
\end{split}
\end{equation}
and from Eq.~(\ref{eqn:entropylocalapp}), we have
\begin{equation}
\label{eqn:s3simm}
\begin{split}s_2(R)=-&\frac{4\pi\rho}{2}\int_{0}^R\biggl( {g}^{(2)}(r)\text{ln} {g}^{(2)}(r)-{g}^{(2)}(r)+1\biggr)r^2\text{d}r\\
\medskip
\medskip
s_3(R)=-&\frac{8\pi^2\rho^2}{6}\int_{0}^R\int_{0}^R\int_{0}^R \biggl[ \\
&{g}^{(3)}(r,s,t)\text{ln}\biggl( \frac{{g}^{(3)}(r,s,t)}{ {g}^{(2)}(r) {g}^{(2)}(s) {g}^{(2)}(t)}\biggr) \\
- &{g}^{(3)}(r,s,t)- {g}^{(2)}(r)- {g}^{(2)}(s) -{g}^{(2)}(t) \\
+&{g}^{(2)}(r){g}^{(2)}(s)+{g}^{(2)}(r){g}^{(2)}(t)+{g}^{(2)}(s){g}^{(2)}(t)\\
+&1\biggr]rst\text{d}r\text{d}s\text{d}t.
\end{split}
\end{equation}
As $R$ increases, the correlation between particles tends to zero and both $s_2$ and $s_3$ converge to their asymptotic values. Our goal is to estimate
\begin{equation*}
\begin{split}
&\lim_{R\to\infty}s_2(R)\\
&\lim_{R\to\infty}s_3(R).
\end{split}
\end{equation*}

Distribution functions are approximated with a histogram with bin size
$\Delta$. During computer simulations, histograms are periodically
updated by counting the number of realizations occurring in each bin,
i.e. all pairs and triplets of particle distances which lay within each
bin volume. Each value is then normalized by the distribution of a
perfect gas in the same macroscopic condition \cite{muller}. These can
be expressed as
\begin{equation}
\label{eqn:entropylocalapp3}
\begin{split}
{g}^{(2)}_i&=\frac{\left \langle N_{i} \right \rangle}{\rho N V^{(2)}_{i}}\\
{g}^{(3)}_{i,j,k}&=\frac{\left \langle N_{i,j,k} \right \rangle}{\rho^2 N V^{(3)}_{i,j,k}}
\end{split}
\end{equation}
\medskip
with 
\begin{equation}
\begin{split}
V^{(2)}_{i}&=\frac{4\pi}{3}\int_{(i-1)\Delta}^{i\Delta} r^2\text{d} r   \\
\\
V^{(3)}_{i,j,k}&=8\pi^2\int_{(i-1)\Delta}^{i\Delta}\int_{(j-1)\Delta}^{j\Delta}\int_{(k-1)\Delta}^{k\Delta}rst\text{d} r\text{d}s\text{d} t.
\end{split}
\end{equation}

Due to the triangle inequality, the domain of the three-particle distribution function is delimited by the planes
\begin{equation}
\label{eqn:entropylocalapp2}
\begin{split}
r&=s+t\\
s&=r+t\\
t&=r+s.\\
\end{split}
\end{equation}
	
Thus, bins that are crossing these borders have a smaller volume, as
they are partially occupied by the domain. We call $R,S,T$ the integer
coordinates of a generic bin, along the directions $r,s,t$ respectively.
A generic triplet $R,S,T$ (with $R,S,T=1,2,3,...$) denotes the region
$[(R-1)\Delta,R\Delta)\times[(s-1)\Delta,S\Delta)\times[(T-1)\Delta,T\Delta)$,
and the triangle inequality implies that the three discrete indices
satisfy
\begin{equation}
\label{eqn:entropylocalapp4}
\begin{split}
R\le S+T\\
S\le R+T\\
T\le R+S.\\
\end{split}
\end{equation}

If we limit our calculation to the region
$[0,R_{max})\times[0,R_{max})\times[0,R_{max})$ the domain is symmetric
by any permutation of indices, and we can focus our analysis on the
region $R\ge S\ge T$, where the triangle inequality takes the form $T\ge
R-S$. In this framework, any bin for which $T\ge R-S+2$ is fully
immersed in the domain of the three particle function, and its volume is 
\begin{equation}
\begin{split}
V^{(3)}_{R,S,T}=8\pi^2\int_{(R-1)\Delta}^{R\Delta}\int_{(S-1)\Delta}^{S\Delta}\int_{(T-1)\Delta}^{T\Delta}rst\text{d} r\text{d}s\text{d} t\\
=8 \pi^2 (R-1/2)(S-1/2)(T-1/2)\Delta^6.
\end{split}
\end{equation}

The border bins ($R-S\le T<R-S+2$), exhibit a smaller volume. Particularly, we can distinguish the case where a bin is crossing one, two or three planes respectively: 
\begin{itemize}
\item one plane: $T=R-S$, $T=R-S+1$, with $R\ne S$, $R\ne 1$, $S\ne 1$;
\end{itemize}
\begin{itemize}
\item two planes: $T=1$, with  $R= S$;
\end{itemize}
\begin{itemize}
\item three planes: $T=R=S=1$.
\end{itemize}

Each distinct case has to be computed separately, i.e. no general
expression, valid for each possible case, can be found. Besides this,
each volume must be further decomposed in smaller elementary volumes. As
an example, the case of three plane intersections must be computed via 4
different integrals: 
\begin{equation}
\begin{split}
V_{1,1,1}=8\pi^2\biggl[&\int_{0}^{\Delta/2}\int_{0}^{r}\int_{r-s}^{r+s}rst\text{d} r\text{d}s\text{d} t\\
+&\int_{\Delta/2}^{\Delta}\int_{0}^{\Delta-r}\int_{r-s}^{r+s}rst\text{d} r\text{d}s\text{d} t\\
+&\int_{\Delta/2}^{\Delta}\int_{\Delta-r}^{\Delta/2}\int_{r-s}^{\Delta}rst\text{d} r\text{d}s\text{d} t\\
+&\int_{\Delta/2}^{\Delta}\int_{\Delta/2}^{r}\int_{r-s}^{\Delta}rst\text{d} r\text{d}s\text{d} t\biggr].
\end{split}
\end{equation}

We do not show the explicit formula of each case, as it has been
provided in a work of Baranyai and Evans\cite{barevans2}. It is clear,
however, that this direct method is tricky, and we expect that it might
be extremely challenging to extend this approach to the computation of
higher order correlation functions, regardless of the other technical
difficulties related to such a calculation.

\subsection{\label{sec:newmeth}Proposed new dimensionless method}

\begin{figure}
\centering
\includegraphics[width=\columnwidth]{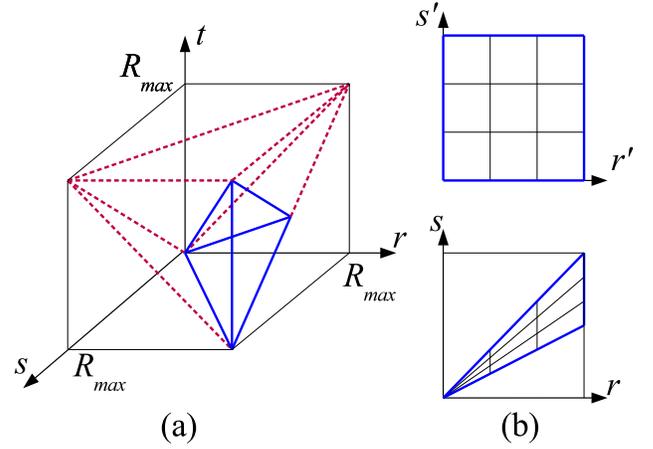}
\caption{(a): The three-particle function's domain (red borders,
dotted), and the region chosen for the dimensionless method (blue
borders, solid). (b) Cubic grid in the dimensionless space for the first
two distances , $r',s'$ and the corresponding grid in the original
space.}
\label{fig:domain}
\end{figure}

We now show a more direct method to discretize the space of the three
distances, such that no correction of border bins is required.
Exploiting the symmetry of $g^{(3)}$ by any permutation of particles,
e.g.
\begin{equation}
\label{eqn:simmetry}
{g}^{(3)}(\textbf{r}_1,\textbf{r}_2,\textbf{r}_3)={g}^{(3)}(\textbf{r}_3,\textbf{r}_2,\textbf{r}_1)\Rightarrow g^{(3)}(r,s,t)=g^{(3)}(t,s,r)
\end{equation}
we can restrict the volume of interest to the region $r\ge s\ge t$, and
combining it with the triangle inequality, we get
\begin{equation}
r \in [0,R_{max}],\;\;\;\;\;\; s\in\biggl[\frac{r}{2},r\biggr] ,\;\;\;\;\;\;t\in[r-s,s]
\end{equation}
where $R_{max}$ is the maximum particle separation distance chosen for
the calculation.  The three distances can now be transformed into
dimensionless counterparts 
\begin{equation}
\label{eqn:dimensionless}
\begin{alignedat}{2}
r'&=\frac{r}{R_{max}}                                         \quad \quad && r'\in[0,1]\\
\\
s'&=\frac{2}{r}\biggl(s-\frac{r}{2}\biggr)   \quad \quad && s'\in[0,1] \\
\\
t'&=\frac{t-r+s}{2s-r}                                \quad\quad && t'\in[0,1]. \\  
\end{alignedat}
\end{equation}
The new domain is trivially cubic (see Figure~\ref{fig:domain}) and so
perfectly fitted by a cubic bin grid. The elementary volume can be
expressed by 
\begin{equation}
\begin{split}
dv(r,s,t)=&rst\text{d} r\text{d}s\text{d} t=\\\
=&\frac{{R^6}_{max}}{4}{(r')}^5s'(s'+1)\biggl(t's'+\frac{1-s'}{2}\biggr)\text{d} r'\text{d}s'\text{d} t'.
\end{split}
\end{equation}
The three-particle correlation function, $g^{(3)}$, as well as $s_3$ can
now be computed more straightforwardly. However, the main inconvenience
of this method is the emergence of an unavoidable mismatch between the
grid used for $g^{(2)}$ and the one used for $g^{(3)}$. In the
computation of $s_3$, the two-particle distribution function is
approximated as constant within each bin of the histogram of $g^{(3)}$,
equal to its value in the barycentre of the bin.

In this work, we computed the two and three-particle entropy of
Weeks-Chandler-Andersen (WCA) particles  in different conditions.  The
WCA potential is defined as \cite{wca}
\begin{equation}
\Phi_{WCA}(r)=\begin{dcases}4\epsilon \biggl[{\biggl(\frac{\sigma}{r}\biggr)}^{12}-{\biggl(\frac{\sigma}{r}\biggr)}^{6}\biggr]+\epsilon & \mbox{if }r\le 2^{\frac{1}{6}}\sigma \\ 0 & \mbox{if }r> 2^{\frac{1}{6}}\sigma
\end{dcases}
\end{equation}
with $r$ the inter-particle distance, $\sigma$ the diameter of the
particles and $\epsilon$ the potential well depth.  We selected a case
of low density ($0.3$), intermediate ($0.7$) and high density ($0.92$),
at reduced temperature of $T=1.15$,and various states close to the
freezing line\cite{wcaphase,WCAphase2} ($\rho=0.92$, $T=0.75$; $\rho=1$,
$T=1.5$; $\rho=1$, $T=2$). We analyzed the numerical instability of the
three-particle entropy and we compared the standard method and the new
dimensionless one. 

\section{\label{sec:compdet}Computational details}

We performed molecular dynamics simulations for systems of 6750 WCA
particles with diameter $\sigma$ and mass $m$ at different states.  The
system size, $L$, was varied based on the density from a minimum of
$L=19.43\sigma$ to a maximum of $L=28.31\sigma$. All variables from now
on are provided in reduced form, and $\sigma=1$, $\epsilon=1$, $m=1$. We
used a fourth order Gear predictor-corrector scheme \cite{alltild} to
integrate the equations of motion, with reduced time step $\Delta
t=0.001$. All systems were thermostatted with a Gaussian isokinetic
thermostat. Periodically, all pairs and triplets of particles which laid
within a distance of $R_{max}=8.6362\sigma$ from each other were used to
update the two and three-particle histograms. The frequency of the
sampling was chosen such that the total simulation time was equally
distributed in the three operations consisting of, respectively,
updating positions and momenta via the integration of the equations of
motion, updating $g^{(2)}$, and updating $g^{(3)}$. Since a single
update of $g^{(2)}$ is computationally faster than an update of
$g^{(3)}$, the frequency of the two updates was different. The number of
time steps between two updates was about $70$ for the two-particle
function and $10^5$ for the three-particle function. These values are
however merely representative, as they were affected by the density of
the system. 

Particles were started in a cubic lattice, and the momenta were randomly
generated and rescaled in order to have zero net momentum along each
direction and the desired initial temperature. An initial warm-up of
$10^5$ time steps was used to allow the system to relax before
commencing the sampling of the distribution functions.  A uniform grid
in the dimensionless space entails that the grid gets coarser as the
separation distance between particles increases (cf.
Figure~\ref{fig:domain}). In order to guarantee a good match between the
two and three-particle grid, the two-particle function was discretized
with a much finer grid than $g^{(3)}$. The three-particle histogram was
created using a grid of $300$ bins per side. As a reference, we use the
bin size along the first distance, $r'$, as it is only rescaled by a
factor of $R_{max}$ (cf. Eq.~(\ref{eqn:dimensionless})). In the real
space, bin size $\Delta_b$ is $R_{max}/300\simeq2.88\times
10^{-2}\sigma$. The two-particle function was discretized with a grid of
$3\times10^6$ bins and thus with a bin size of $\Delta=10^{-4}\Delta_b$.
Each $g^{(3)}$ was obtained via $\simeq 850$ samples, for a total amount
of $\simeq 2.7\times10^{12}$ triplets sampled from each run, resulting
in an average number of realizations per bin $\bar{N}_b\simeq10^5$.
About $7.5\times 10^5$ samples and $\simeq 1.2\times10^{13}$ total pairs
were sampled for $g^{(2)}$, with an average number of realizations per
bin of $\simeq4\times10^6$. 

The selection of the triplets of particles requires some consideration.
For a generic particle $i$, all neighbor particles, i.e. closer than
$R_{max}$, are identified. This procedure guarantees that, given two
neighbors $j$, $k$, the inter-particle distances $r_{ij}$ $r_{ik}$,
$r_{jk}$ are in the following ranges
\begin{equation}
\begin{split}
&0\le r_{ij}< R_{max}\\
&0\le r_{ik}< R_{max}\\
&0\le r_{jk}< 2R_{max}\\
\end{split}
\end{equation}
The constraint of avoiding self interaction requires every distance to
be smaller than $L$. On the other hand, the presence of any distance in
the range $[L/2,L)$ entails that more than one set of distances may be
found for the same triplet of particles, due to interaction with
particle images. It is usually preferred to set $R_{max}=L/4$ and avoid
any possible multiple interaction\cite{maxdist}. In a previous
work\cite{barevans2} the two ranges $R_{max}=L/2$ and $R_{max}=L/4$ were
tested, and no statistical difference was found between them.  We note
that with such a scheme, the distances are not symmetric by a
permutation of particles. This gives rise to two further problems: the
domain of $g^{(3)}$ is no longer cubic and symmetric, as depicted in
Sec.~\ref{sec:newmeth}, and the calculation of the three distances must
be performed for any permutation of $i$, $j$, $k$, with an increment of
the number of calculations of 6 times. In this work, we preferred to set
the range of each distance
\begin{equation}
\begin{split}
&0\le r_{ij}< R_{max}\le L/2\\
&0\le r_{ik}< R_{max}\le L/2\\
&0\le r_{jk}< R_{max}\le L/2.
\end{split}
\end{equation}
This constraint entails some advantages:
\begin{itemize}
\item any permutation of $i$, $j$, $k$ locates the same triplet of
distance, making the update of $g^{(3)}$ 6 times faster, and allowing to
reduce the domain in the way previously described, making the histogram
of $g^{(3)}$ 6 times smaller. 
\item $R_{max}$ can be set up to $L/2$ avoiding a too cautious constraint of $L/4$.
\end{itemize}

Once a triplet of particles was obtained, the three distances were
sorted and rescaled with the formula provided in
Eq.~(\ref{eqn:dimensionless}), and the corresponding bin was updated.  

The simulations were performed on the Swinburne supercomputer cluster
OzSTAR, using multiple central processing unit (CPU) Intel Gold 6140
processors. On these CPUs, each simulation was about 156 hours long,
each time step was performed in $\simeq 1.8\times10^{-3}$s, each update
of $g^{(2)}$ required about $1.2\times10^{-1}$s, and each update of
$g^{(3)}$ about $186$s.

\section{\label{sec:results}Results and discussion}

\begin{figure}
\centering
\includegraphics[width=\columnwidth]{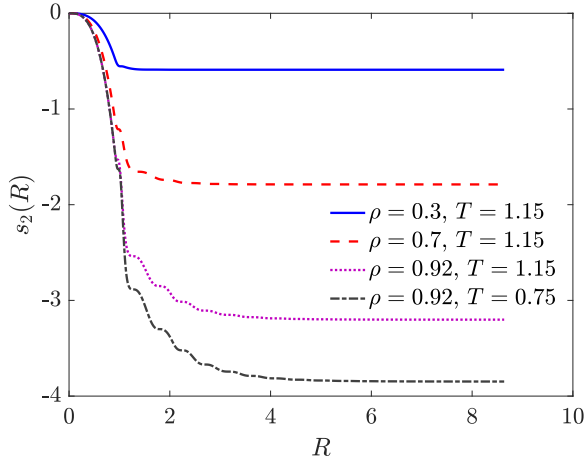}
\caption{Two-particle entropy $s_2$ for different conditions.}
\label{fig:s2}
\end{figure}

\begin{figure}
\centering
\includegraphics[width=\columnwidth]{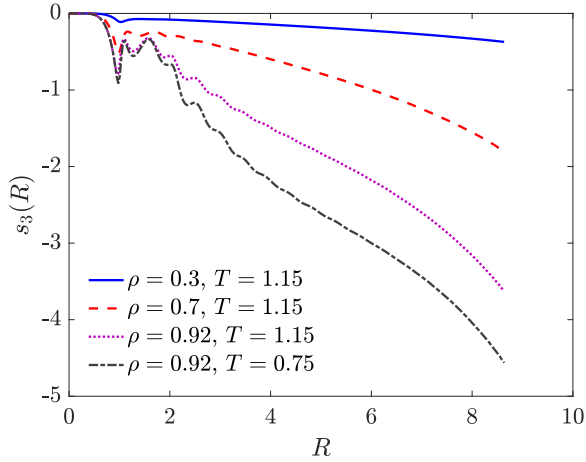}
\caption{Three-particle entropy $s_3$ for different conditions.}
\label{fig:s3}
\end{figure}

Results in Figures~\ref{fig:s2}-\ref{fig:s3} show the two and
three-particle entropy computed at different conditions of density and
temperature. The three-particle entropy is affected by the numerical
drift. Previous analysis \cite{barevans2} ascribed the drift mainly to
statistical fluctuations due to the small number of samples, and to a
lesser extent, to the grid (i.e. bin size) used to histogram $g^{(3)}$.
We found that, when the new dimensionless method is used, both the small
number of samples and the finite bin size $\Delta$ have a key role in
the promotion of the drift, as we will now discuss.

\subsection{\label{sec:noise}Effect of statistical noise}

\begin{figure}
\centering
\begin{subfigure}{\columnwidth}
\includegraphics[width=\textwidth]{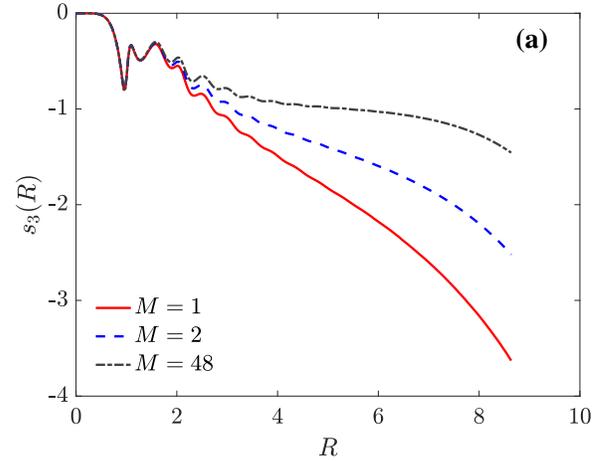}
\label{fig:noise1}
\end{subfigure}

\begin{subfigure}{\columnwidth}
\includegraphics[width=\textwidth]{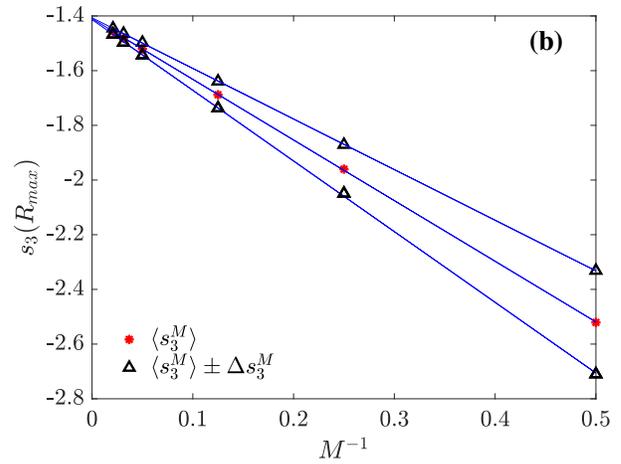}
\label{fig:noise3}
\end{subfigure}
\caption{\textbf{(a)} Magnitude of the drift for different numbers of
samples. \textbf{(b)} Convergence of systematic effect (sample average,
red points) and statistical fluctuations around the sample average,
depicted via the standard deviation $\Delta s_3$ (black triangles), with
respect to the number of samples. Blue lines represent the linear
interpolation of both effects. State point $\rho=0.92$,
$T=1.15$.}\label{fig:noise}
\end{figure}

Numerical error induced by a small number of samples has a twofold
aspect: due to the non linear expression of $s_3$, statistical
fluctuation of $g^{(3)}$ around its true value results in both
statistical and systematic error (i.e. drift) in the three-particle
entropy.  Increasing the number of samples reduces the error,  however,
an estimate of the residual drift and statistical fluctuations is
necessary, as well as a method to extrapolate the distribution of the
three-particle entropy in the infinite sample limit.  Our goal is to
generate the statistical distribution of the three-particle entropy for
different levels of noise (i.e. equivalently, for different number of
samples used to compute the two and three-particle correlation
functions).  A set of 48 two and three-particle distributions was
obtained from the same number of independent simulations. Each
distribution had the number of samples and bin size described in
Sec.~\ref{sec:compdet}. Merging together a certain number $M$ of these
distribution functions allows us to increase the number of samples used
to compute $s_3$, roughly of the order $M$. We define
\begin{equation}
s_3(R)^M=s_3(\langle g^{(2)}\rangle_M,\langle g^{(3)}\rangle_M)
\end{equation}
the statistical distribution of the entropy computed via $M$ two and
three-particle functions merged together.  We selected a set of values
$M$ ($M=2,4,8,20,32,48$) and for each value, the statistical
distribution of $s_3$ was approximated via 200 different computation of
$s_3$: each $s_3$ was computed by merging $M$ two and three-particle
functions, randomly extracted with repetitions from the original 48
distributions. Results for the state point $\rho=0.92$, $T=1.15$ are
shown in Figure~\ref{fig:noise}. It is clear that both the systematic
and the statistical fluctuations converge with a law $\propto
M^{-1}\propto {N_s}^{-1}$, with $N_s$ the number of samples used to
generate $g^{(2)}$ and $g^{(3)}$. These results are in good agreement
with previous works\cite{barevans2}.

This procedure allowed us to determine the law of the convergence of
$s_3$ with respect to the statistical noise in $g^{(2)}$ and $g^{(3)}$.
However, these data cannot be used to extrapolate the infinite sample
limit, as they are highly correlated. The extrapolation would not differ
substantially from the direct computation of $s_3$ when all $48$
distributions are merged together without repetitions, and a linear
drift is still well detectable even in the extrapolated function.  In
order to remove the bias induced by the inter-dependence of the data, a
different procedure was used: with a proper choice of the values $M$, it
is possible to generate a sequence of $s_3$ with different levels of
noise in which each one of the original $48$ distributions appears only
once. A possible choice could be $M=1,2,4,8,33$. As we see, the sum of
all $M$ is 48, so no repeated distributions appear in the sequence. The
$48$ distributions are merged together with the rule previously
described. In this procedure, one single $s_3$ is associated to each
value of $M$. As an example, calling $g^{(n)}_i$ the $i$-th distribution
function over the original 48, the set $\{{s_3}^M\}$ may be composed by
$5$ different $s_3$ computed in the following way:
\begin{equation}
\label{eqn:1}
\begin{split}
{s_3}^1&=s_3( g^{(n)}_1)\\
{s_3}^2&=s_3(\langle g^{(n)}_{2,3}\rangle)\\
{s_3}^4&=s_3(\langle g^{(n)}_{4,...,7}\rangle)\\
{s_3}^8&=s_3(\langle g^{(n)}_{8,...,15}\rangle)\\
{s_3}^{33}&=s_3(\langle g^{(n)}_{16,...,48}\rangle)\\
\end{split}
\end{equation}

\begin{figure}
\centering
\includegraphics[width=\columnwidth]{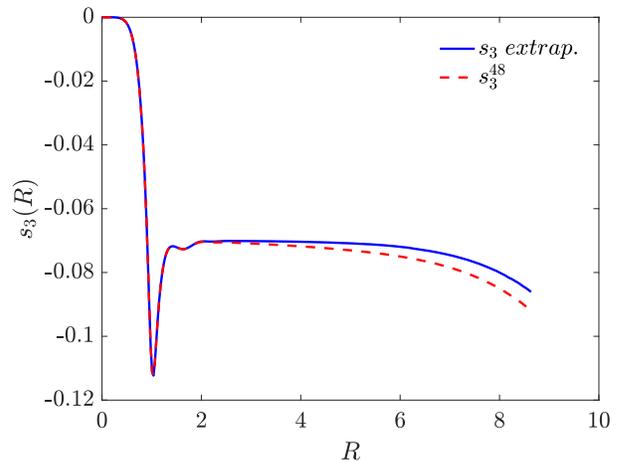}
\caption{Three-particle entropy $s_3$ for $\rho=0.3$, $T=1.15$ obtained via the sum of all 48 distributions and extrapolated to the zero noise limit.}
\label{fig:extravsnoextrap}
\end{figure}

\begin{figure}
\centering
\includegraphics[width=\columnwidth]{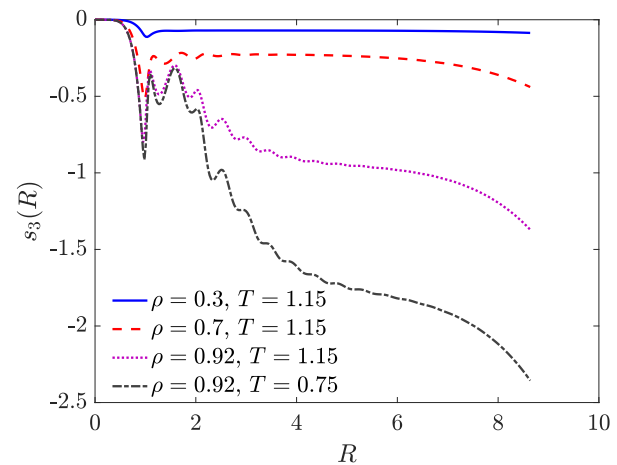}
\caption{Three-particle entropy $s_3$ for different states, extrapolated to zero noise limit.}
\label{fig:s3nonoise}
\end{figure}

This sequence was used to extrapolate the infinite sample limit with the
law previously derived. The results are unbiased, as the data are
independent,  but they may be still affected by the order in which the
$48$ distributions appear in the sequence of Eq.~(\ref{eqn:1}). In order
to remove this effect, we generated the statistical distribution of the
extrapolated value via 600 different computations obtained permuting the
order of the $48$ distributions in Eq.~(\ref{eqn:1}). The sample average
has then been selected as the estimate of $s_3$ in the infinite sample
limit. Results are shown in
Figures~\ref{fig:extravsnoextrap}-\ref{fig:s3nonoise}. It is evident
that an extrapolation is necessary, particularly at low densities, as
the small value of the true entropy and a smaller drift induced by the
grid size makes the noise effect the primal source of error in the
estimate of $s_3$. At higher densities, grid size effect is largely
increased, and the noise has a secondary role in the drift.  The
standard deviation of the extrapolated $s_3$ is roughly constant for all
different systems, and it is up to 0.02 at $R=R_{max}$. At that
distance, however, accuracy of the extrapolation is nullified by the
grid size effect. For shorter distances it becomes almost negligible,
being about 0.005 at $R=R_{max}/2$.

\subsection{\label{sec:grid}Effect of the discretization}

At low and intermediate densities the grid size effect is essentially
negligible, since $s_3$ reaches its asymptotic value well before the
drift starts to be relevant. For $\rho=0.3$ and $T=1.15$ the convergence
is very fast, and $s_3$ can be considered converged  at $R=3\sigma$; at
intermediate densities, the inter-particle distance of convergence is
about $4\sigma$. In these cases, the grid size effect can be ignored, as
$s_3$ exhibits a clear plateau with neither oscillations nor drift. At
high densities, oscillations have not yet ceased when numerical
instability starts to become relevant.  A finer grid is able to reduce
the magnitude of the drift; however, this would require much larger
computational resources for storing the histograms, whose size scales
with the third power of the number of bins per side, and much longer
runs, to have an acceptable level of noise. 

\begin{figure}
\centering
\includegraphics[width=\columnwidth]{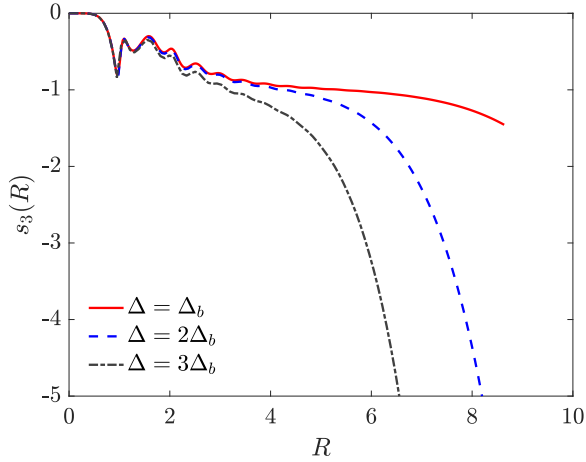}
\caption{Grid size effect on the three-particle entropy ($\rho=0.92$, $T=1.15$), with $\Delta_b = 2.88 \times 10^{-2}$ in reduced units.}
\label{fig:gridfinal}
\end{figure}

\begin{figure}
\centering
\includegraphics[width=\columnwidth]{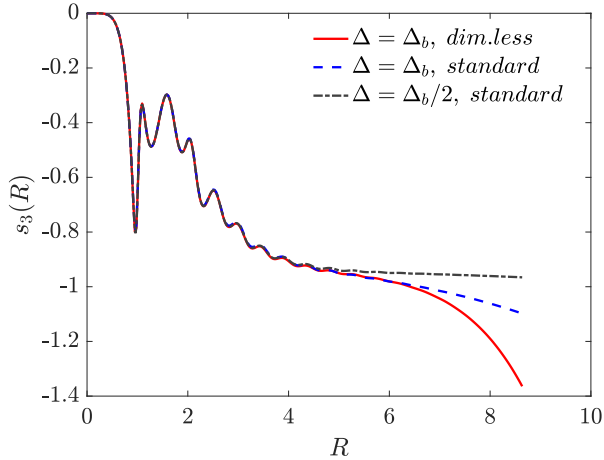}
\caption{Grid effect on standard and dimensionless method ($\rho=0.92$, $T=1.15$). $\Delta_b = 2.88 \times 10^{-2}$ in reduced units.}
\label{fig:griddimadim}
\end{figure}

In Figure~\ref{fig:gridfinal} we see that the magnitude of the drift is
extremely sensitive to the grid size. This phenomenon is particularly
severe for high densities. Figure~\ref{fig:gridfinal} shows the
magnitude of the drift for different bin sizes. The convergence is clear
as bin size decreases, however, no simple law has been found able to fit
the data with accuracy acceptable to have a reliable correction of the
drift. We detected at least two different kinds of error dependent on
the bin size: the numerical instability, resulting in the drift, and the
error caused by the smoothing of the structure of the correlation
functions, induced by the histogram approximation. When the bin size is
increased, the resulting distribution functions are smoother, and closer
to those of a perfect gas. This effect manifests itself in a systematic
\textit{under-estimation} of the magnitude of $s_2$ and $s_3$. The
second type of error relies on the intimate structure of the correlation
functions, and is thus essentially unpredictable without a new
theoretical insight into the correlation functions. The interaction
between these two effects makes the correction of the drift quite
challenging.  At this stage the only robust correction of both effects
can be achieved by simply choosing a grid fine enough to reduce both
errors to a tolerable level. From Figure~\ref{fig:griddimadim} we see
that the drift is generally less severe when the standard method (rather
than the dimensionless one) is used, the baseline bin size $\Delta_b$
being equal. By exploiting the symmetry of the three-particle function,
the histogram of the standard $g^{(3)}$ can be reduced in size, and thus
a finer grid can be used. A grid of $\Delta=\Delta_b/2$ is able to
reduce the drift to a tolerable level, and $s_3$ shows excellent
convergence.   

\begin{figure}
\centering
\includegraphics[width=\columnwidth]{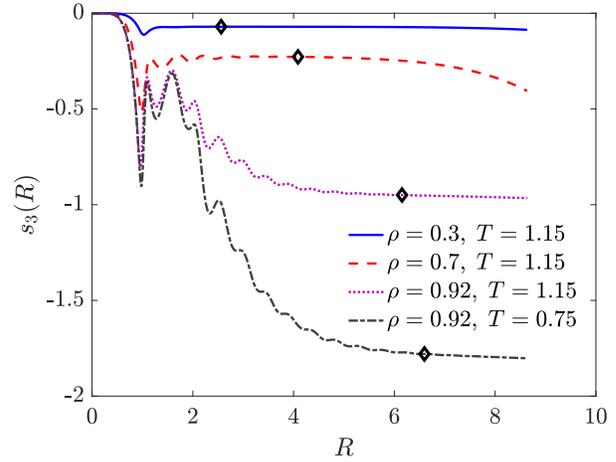}
\caption{$s_3$ for different state points, and its convergence distance depicted by the
diamond points.}
\label{fig:convradius}
\end{figure}

\begin{figure}
\centering
\includegraphics[width=\columnwidth]{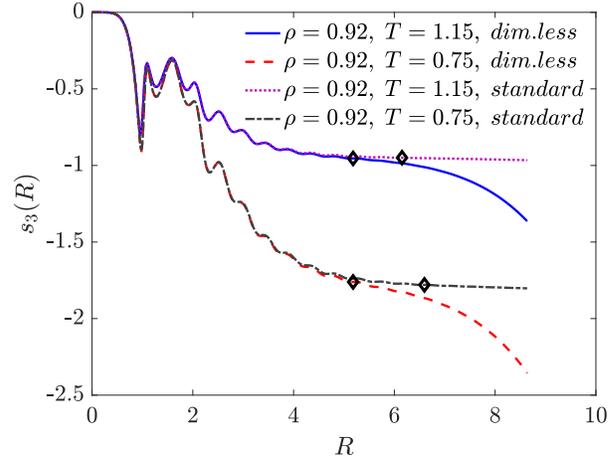}
\caption{$s_3$ and convergence distance for standard and dimensionless method.}
\label{fig:convradiusdimadim}
\end{figure}

\subsection{\label{sec:final}Final estimate}

The presence of numerical instability implies that the convergence of
$s_3$ can never be reached, thus, a procedure to select its asymptotic
value is necessary. The cases of low and intermediate density suggests
that when oscillations cease the three-particle entropy is fully
converged (Figure~\ref{fig:convradius}). At high density, however, the
effect of the drift becomes clearly apparent before oscillations become
negligible. The best compromise between these two opposite effects was
found in the point $R_{conv}$ in which residual oscillations and
numerical instability have the same magnitude. This distance was
selected as the separation distance at which the entropy can be
considered converged. The region of incomplete convergence $R<R_{conv}$
is characterized by the presence of stationary points (mimina and
maxima), regularly distributed along $R$. When the drift dominates the
residual oscillations, at $R>R_{conv}$, the first derivative of $s_3$
is, on the other hand, systematically negative. The last stationary
point occurring in each $s_3$ was thus selected as $R_{conv}$ and the
asymptotic three-particle entropy was set as $s_3(R_{conv})$.
Figure~\ref{fig:convradiusdimadim} shows that the numerical instability
affecting the dimensionless method with grid size $\Delta_b$ leads to a
significantly incorrect estimate of $R_{conv}$ (represented by the
diamond symbol), as weak oscillations can still be detected after this
distance. The standard method provides a more reliable estimate of
$R_{conv}$, although, quite encouragingly, the relative difference
between the entropies was about $0.4\%$ at $T=1.15$ and $1\%$ at
$T=0.75$. 

\begin{figure}
\centering
\includegraphics[width=\columnwidth]{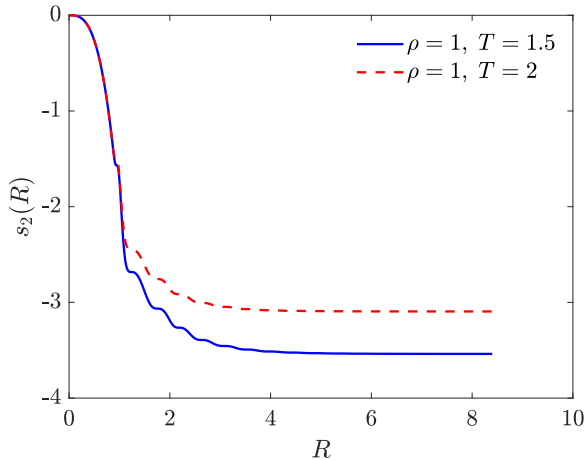}
\caption{$s_2$ for different state points close to the freezing line.}
\label{fig:s2melting}
\end{figure}

\begin{figure}
\centering
\includegraphics[width=\columnwidth]{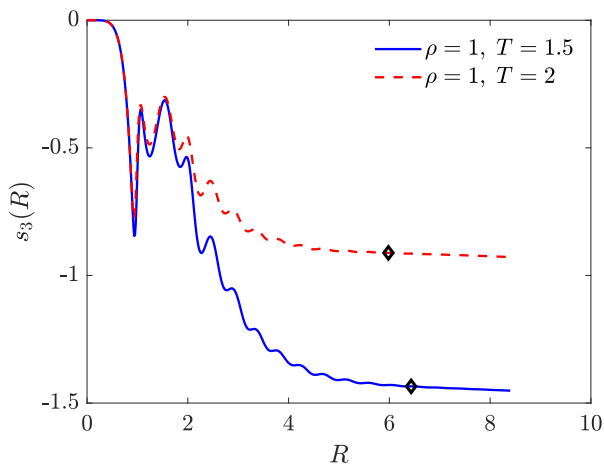}
\caption{$s_3$ for different state points close to the freezing line ($\Delta=\Delta_b/2$).}
\label{fig:s3melting}
\end{figure}

The state points close to the freezing line were investigated via the
standard method, due to their high density. Results are shown in
Figures~\ref{fig:s2melting} and \ref{fig:s3melting}. For reference, the
state points at the solid-liquid phase boundary (coexisting liquid
densities) for $\left(\left\{T\right\},\left\{\rho\right\}\right) =
\left(\left\{0.75,1.15,1.5,2\right\},\left\{0.915,0.98,1.02,1.07\right\}\right)$
are taken from Ahmed and Sadus \cite{WCAphase2}.

Table~\ref{Tab:Tcr} provides the final estimate of the two and
three-particle entropy for all states. The error of the estimate is
expressed via the standard deviation $\Delta s_3$ of the distribution of
the values obtained from the extrapolation to the infinite sample limit
at $R=R_{conv}$ (diamond points in Figure~\ref{fig:convradius},
\ref{fig:s3melting}).  Comparing the results of WCA with Lennards-Jones
particles\cite{barevans2}, we found that the WCA two and three-particle
entropy exhibit the same behavior as the Lennard-Jones entropy, the
speed of convergence (in terms of separation distance $R$) and the
oscillatory behavior being essentially the same. The Lennard-Jones
entropy has, however, a systematically larger magnitude than the WCA
one, in accordance with the attractive component and slower delay of
correlations for the Lennard-Jones potential.  The two and
three-particle entropies were compared with the excess entropy per
particle $s_{ex}$
\begin{equation}
\label{eqn:excessentropy}
\begin{split}
s_{ex}=s-s_{pg}&=\frac{\beta U_{ex}}{N}-\frac{\beta F_{ex}}{N}\\
\frac{\beta U_{ex}}{N}&=\frac{\langle  E_{pot}\rangle}{Nk_B T}\\
\frac{\beta F_{ex}}{N}&=\int_0^{\rho}\frac{Z-1}{\rho'}\text{d}\rho'
\end{split}
\end{equation}
where $ U_{ex}$ and $ F_{ex}$ are the excess internal energy and the
excess Helmholtz free energy. The average potential energy $\langle
E_{pot}\rangle$, as well as the compressibility factor $Z$, were
computed via a WCA equation of state recently found\cite{wcaEOS}.
Table~\ref{Tab:Tcr2} highlights the magnitude of the two and
three-particle entropy with respect to the total excess entropy.

The ratio $s_3/s_2$ remains quite stable up to intermediate densities
($0.119$ at $\rho=0.3$, $0.127$ at $\rho=0.7$) and then increases
dramatically when the system is approaching the freezing point ($0.297$
at $\rho=0.92$, $T=1.15$). Close to the freezing line, $s_3$ is about
$45\%$ of $s_2$ at $T=0.75$, and it then decreases with increasing
temperature (less than $30\%$ of the two-particle entropy at $T=2$).  At
high densities the sum of the first two terms, $s_2$ and $s_3$, exceeds
the total excess entropy, a phenomenon already detected for both
Lennard-Jones and hard spheres systems\cite{barevans2,barevans3}. This
fact entails that the total contribution of higher order terms in the
Green's expansion must be positive. We observe that when the freezing
line is approached, the overestimation by the three-body contribution to
the excess entropy is almost totally nullified by the higher order
terms, and the approximation of $s_{ex}$ with $s_2$ alone provides a
more accurate estimate than $s_2+s_3$ (Table~\ref{Tab:Tcr2}). This
effect is enhanced by low temperatures, with $s_2$ being more than
$97\%$ of the excess entropy at $\rho=0.92$ and $T=0.75$. 

We are aware that the data presented in the work of Ahmed and Sadus
\cite{WCAphase2} places the state point $\rho=0.92$, $T=0.75$ in the
region of fluid-solid coexistence. However, the uncertainty about the
freezing density and temperature is quite large, as pointed out by the
authors themselves when they compare their results with a previous work
(see de Kuijper {\it{et. al.}} \cite{wcaphase}), in which, for instance,
the state point $\rho=0.92$, $T=0.75$ has been found to be still in the
fluid regime. The lack of any apparent anomalous behaviour in our
computation of the two and three-particle entropy suggests that our
results are reliable in the evaluation of the magnitude of such
quantities in proximity to that state point. 

\begin{table*}
\caption{\label{Tab:Tcr}Asymptotic values of two and three-particle entropies at different states. $R_{conv}$ is the convergence distance of $s_3$. $\Delta s_3$ is the standard deviation of the distribution obtained for the zero noise extrapolation at $R=R_{conv}$.}
\begin{ruledtabular}
\begin{tabular}{ccccccccc}
$\rho$ & $T$ & $s_2$ &$s_3$ & $R_{conv}$& $\Delta s_3$ \\ \hline
\\
$0.3$ & $1.15$ & $-0.5900$ &$-0.0700$ & $2.57\sigma$& $1.0\times 10^{-4}$\\
$0.7$ & $1.15$ & $-1.7880$ &$-0.2276$ & $4.09\sigma$& $5.7\times10^{-4}$\\
$0.92$ & $0.75$ &$-3.8480$ &$-1.7798$ & $6.42\sigma$& $7.1\times10^{-4}$\\
$0.92$ & $1.15$ &$-3.2012$ &$-0.9490$ & $6.15\sigma$& $5.2\times10^{-4}$\\
$1$ & $1.5$ &$-3.5383$ &$-1.4346$ & $6.43\sigma$& $4.8\times10^{-4}$\\
$1$ & $2$ &$-3.0945$ &$-0.9115$ & $5.98\sigma$& $4.4\times10^{-4}$\\
\end{tabular}
\end{ruledtabular}
\end{table*}

\begin{table*}
\caption{\label{Tab:Tcr2}Magnitude of $s_2$ and $s_3$ with respect to the total excess entropy computed from the equation of state of Mirzaeinia {\it{et. al.}}\cite{wcaEOS}.}
\begin{ruledtabular}
\begin{tabular}{ccccccccc}
$\rho$ & $T$ & $s_3/s_2$ &$s_{ex}$ & $s_2/s_{ex}$ & $(s_2+s_3)/s_{ex}$ \\ \hline
\\
$0.3$ & $1.15$ & $0.1186$ &$-0.7118$ & $0.8289$ & $0.9272$ \\
$0.7$ & $1.15$ & $0.1273$ &$-2.2033$ & $0.8115$ & $0.9148$ \\
$0.92$ & $0.75$ &$0.4625$ &$-3.9551$ & $0.9729$ & $1.4229$ \\
$0.92$ & $1.15$ &$0.2965$ &$-3.4823$ & $0.9193$ & $1.1918$ \\
$1$ & $1.5$ &$0.4054$ &$-3.7255$ & $0.9498$ & $1.3348$ \\
$1$ & $2$ &$0.2946$ &$-3.4067$ & $0.9084$ & $1.1759$ \\
\end{tabular}
\end{ruledtabular}
\end{table*}


\section{\label{sec:conclusionsl}Conclusions}

We have presented a new method for the computation of both the
three-particle correlation function and the three-particle entropy for a
system of $N$ interacting particles. Although the procedure for the
volume correction is well known in the literature, the new dimensionless
method involves more straightforward calculations, and it is likely to
be more easily generalized to higher order distribution functions, for
example to compute the four-particle entropy.  The dimensionless method
was tested for a WCA fluid at various state points.  The numerical
instability results from both small number of samples and grid size used
to create the histogram for $g^{(3)}$. While the noise effect can be
removed with a proper extrapolation, the drift induced by finite bin
size is related to the basic structure of the two and three-particle
distribution functions, and is thus not easily removed without an
extensive investigation of such properties.  The dimensionless method is
generally more unstable than the traditional one with respect to the
grid size, with the computational cost being roughly equal. For low and
intermediate densities (up to $0.7$), the numerical instability does not
compromise the computation of the three-particle entropy. Higher
densities, however, require either a finer grid, or the use of the
traditional method.  We found that a grid size of $2.88\times
10^{-2}\sigma$ is fine enough to provide a reliable estimate in most of
the cases, with both the standard and dimensionless methods. Coarser
grids should be avoided, as they result in greater drift. Since, at this
stage, the drift cannot be removed, the selection of a proper estimate
for $s_3$ requires an adequate criterion. We found that selecting the
last stationary point provides a simple and deterministic method, and
the best compromise between the error induced by both the numerical
instability and the incomplete convergence.

Close to the freezing line, the relative over-estimation of the total
excess entropy when the three-body term is included is significantly
higher than the underestimation obtained when only the two-body
contribution is taken into account.
\newline

\section*{Acknowledgements}
  N.C. gratefully acknowledges support from the Australian Research Council under the Future Fellowship scheme (project number FT130100972).


%

\end{document}